\documentclass[12pt,preprint]{aastex}

\shorttitle{Spatial Distribution of Interstellar Dust Medium}
\shortauthors{Xiang, Zhang, \& Yao}

\begin{document}

\title{Probing the Spatial Distribution of the Interstellar Dust Medium by High Angular Resolution X-ray Halos of Point Sources}

\author{Jingen Xiang\altaffilmark{1}, Shuang Nan Zhang\altaffilmark{1,2,3,4}}
\affil{\altaffilmark{1}Physics Department and Center for
Astrophysics, Tsinghua University, Beijing, 100084, China}
\affil{\altaffilmark{2}Physics Department, University of Alabama
in Huntsville, Huntsville, AL35899, USA}
\affil{\altaffilmark{3}Space Science Laboratory, NASA Marshall
Space Flight Center, SD50, Huntsville, AL35812, USA}
\affil{\altaffilmark{4}Institute of High Energy Physics, Chinese
Academy of Sciences, Beijing, China}
\email{xjg01@mails.tsinghua.edu.cn; zhangsn@tsinghua.edu.cn}
\and
\author{Yangsen Yao}
\affil{Department of Astronomy, University of Massachusetts, Amherst, MA 01003}



\begin{abstract}
We studied the X-ray dust scattering halos around 17 bright X-ray point sources
using {\it Chandra} data. All sources were observed with the Advanced CCD Imaging
Spectrometer and High-Energy Transmission Grating in {\it Continuous Clocking Mode}
(CC-mode) or {\it Timed Exposure Mode} (TE-mode). We use an  iterative method to
resolve the halos at small scattering angles from the zeroth order data in CC-mode or
the first order data in TE-mode which is not or less piled-up. Using the interstellar
grain models of Weingartner $\&$ Draine (2001, WD01) and  Mathis, Rumpl $\&$ Nordsieck
(1977, MRN ) to fit the halo profiles, we get the hydrogen column densities and the
spatial distributions of the scattering dust grains along the line of sights (LOS) to
these sources. We find that the scattering dust density very close to these sources is
much higher than the normal interstellar medium. For X-ray pulsars GX 301-2 and Vela
X-1 with companions of strong stellar winds, the X-ray absorption column densities are
much higher than the derived scattering column densities, because of the dense media
around the X-ray sources produce extremely small angle scatterings which are
indistinguishable from the point sources even with \textit{Chandra}'s angular
resolution. For LMC X-1, most of the scattering and absorption occur in Large
Magellanic Cloud, rather than in the Milky Way. From the obtained X-rays spectra, the
cold gas absorption and thus the equivalent hydrogen column is determined. We have
obtained the linear correlation between $N_{H}$ derived from spectral fits and the one
derived from the grain models WD01 and MRN (except for GX 301-2 and Vela X-1):
$N_{H,WD01} = (0.720\pm0.009) \times N_{H,abs} + (0.051\pm0.013)$ and $N_{N, MRN} =
(1.156\pm0.016) \times N_{H,abs} + (0.062\pm0.024)$ in the units $10^{22}$ cm$^{-2}$.
High angular resolution X-ray dust scattering halos offer an excellent potential for
probing the spatial distributions of interstellar dust medium.
\end{abstract}

\keywords{dust --- scattering --- X-rays: ISM}

\section{Introduction}

X-rays are absorbed and scattered by dust grains when they travel through the
interstellar medium. The scattering within small angles results in an X-ray ``halo''. The halo properties are significantly affected by the energy of radiation, the optical depth of the scattering, the grain size distributions and compositions, and the spatial distribution of dust along the line of sight (LOS) (Overbeck 1965, Martin 1970, Catura 1983). Therefore analyzing the the X-ray halo properties is an important tool to study the interstellar grains, which play a central role in the astrophysical study of the interstellar medium, such as the thermodynamics and chemistry of the gas and the dynamics of star formation (Draine, 2003).

Overbeck (1965) was the first to discuss the existence of dust scattering halos.
Mathis, Rumpl $\&$ Nordsieck (1977) derived the dust grain size distributions assuming
a silicate-graphite dust mixture. Rolf (1983) first observationally confirmed the
existence of halos using the the data of GX 339-4 observed with the \textit{Einstein
Observatory}. Predehl $\&$ Schmitt (1995) studied the X-ray halos by systematically
examining 25 point sources and 4 supernova remnants with \textit{ROSAT} observations.
They found a good correlation between the simultaneously measured dust and hydrogen
column densities, as well as a correlation between the visual extinction and the X-ray
derived dust scattering optical depth. Because of the poor angular resolution of the
\textit{ROSAT}, the halos in small angles, e.g. less than 50 arcsec, cannot be
determined accurately; the flat angular distribution of X-ray halos in small angles
determined with \textit{ROSAT} may be due to the underestimation of the halo intensity
in the core region. Draine $\&$ Tan (2003) analyzed the halos of X-ray Nova V1974 Cygni
1992 with the \textit{ROSAT} observations to test the interstellar dust model WD01 and
found that the WD01 model is very consistent with the observed X-ray halos. With the
high angular resolution, good energy resolution and broad energy band, the
\textit{Chandra} ACIS is so far the best instrument for studying the X-ray halos. Tan
$\&$ Draine (2004) considered arcsecond scale X-ray scattering halos from Galactic
Center sources with the \textit{Chandra} observations and showed the halo profile
changed with the spatial distribution of the dust that is close to the source. Smith,
Edgar $\&$ Shafer (2002) presented the observations of the X-ray halo around the Low
Mass X-ray Binary (LMXB) GX 13+1. However, the direct images of bright sources obtained
with Advanced CCD Imaging Spectrometer (ACIS)\footnote{http://cxc.harvard.edu/proposer/POG/html/ACIS.html} usually suffer from
severe pileup. They therefore investigated only the halo in off-axis angles greater
than $50^{''}$ to avoid any possible pileup contamination.

Making use of the assumption that the real halo could be an isotropic image, Yao
\emph{et al} (2003, here after Paper I) reported the reconstruction of the images of
X-ray halos in small angles from the zeroth order data in CC-mode or the first order
data in TE-mode. Xiang, Zhang $\&$ Yao (2004, here after Paper II) improved the method
by solving the equations with iterative method and using the observational data to
generate the \textit{Chandra} Point Spread Function (PSF). They applied the method to
the bright sources Cygnus X-1 and Cygnus X-3 and found that the halos in small angles
are not flat, which is different from the result of Predehl $\&$ Schmitt (1995). Since
the spatial distribution of the dust grains, especially very close to the source, is
sensitive to the core region of an X-ray halo (Mathis $\&$ Lee 1991), it is necessary
to study systematically the halos in small angles for many X-ray point sources.

In this paper, we apply our method described in Paper II, to systematically study the
X-ray halos of 17 X-ray point sources. We resolve the point
source halo in very small angles from the zeroth order data in CC-mode or the grating
data in TE-mode with \textit{Chandra}, and
use the different interstellar grain models to fit the halos and
get the spatial distribution of the interstellar dust medium (\S2).
We discuss our results in \S3 and present our conclusions \S4.

\section{Observations and data reduction}
In this study, we utilized 31 {\sl Chandra} ACIS High Energy Transmission Grating
(HETG)\footnote{http://space.mit.edu/HETG/index.html}
observations on total 19 sources and all the data are available in the
archive upto November 2004. All these sources (except for PKS
2155-304 and Her X-1) are bright X-rays sources and have been observed for at least
8.9 ks, therefore there are enough photons (about $\geq$ 0.5 million counts)
for us to resolve the X-ray halos. Both PKS 2155-304 and Her X-1 are very
lightly-absorbed X-ray sources with
the column density $N_{H}<3 \times 10^{20}$ cm$^{-2}$. We use these two sources to
generate the PSF. We list the source properties and the observation logs in Table 1.


\begin{deluxetable}{rrcrrrrrr}
\tabletypesize{\scriptsize} \tablecaption{List of sources (column
1) with the observation IDs (column 2),observation mode (column 3), positions in equatorial (column 4, 5) and galactic coordinates (column 6, 7), the total counts in 1.0-5.0 keV (column 8), and the total exposure time (column 9)} \tablewidth{0pt} \startdata
\tableline \tableline
Source & ObsID & mode & R.A.[2000] & Dec.[2000] & $l^{II}$ & $b^{II}$ & counts & time (ks)\\
\tableline
4U1705-44 & 1923 & TE & $17^{h}08^{m}55^{s}$ &  $-44^{d}06^{'}00^{''}$ & $343.3^{d}$ & $-2.3^{d}$ & 2637040 & 25.3\\
4U1728-16 & 703 & TE & $17^{h}31^{m}44^{s}$ & $-16^{d}57^{'}42^{''}$ & $8.5^{d}$ & $9.0^{d}$ & 2619501 & 20.9\\
Cir X-1 & 1700, 1905 & TE & $15^{h}20^{m}41^{s}$ & $-57^{d}10^{'}01^{''}$ & $322.1^{d}$ & $0.0^{d}$ & 5208035 & 40.3\\
 & 1906, 1907 & & & & & & &\\
Cyg X-1 & 107, 3814 & TE & $19^{h}58^{m}22{s}$ & $+35^{d}12^{'}06{''}$ & $71.3^{d}$ & $3.1^{d}$ & 8225397 & 58.6\\
Cyg X-2 & 1102 & TE & $21^{h}44^{m}41^{s}$ & $+38^{d}19^{'}18^{''}$ & $87.3^{d}$ & $-11.3^{d}$ & 3513440 & 14.6\\
Cyg X-3 & 425 & TE & $20^{h}32^{m}27{s}$ & $+40^{d}57^{'}10{''}$ & $79.8{d}$ & $0.7^{d}$ & 1431689 & 16.2\\
GRS 1915+105 & 660 & TE & $19^{h}15^{m}12{s}$ & $+10^{d}56^{'}44^{''}$ & $45.4^{d}$ & $0.2^{d}$ & 3379930 & 30.6\\
GX 13+1 & 2708 & TE & $18^{h}14^{m}32{s}$ & $-17^{d}09^{'}27^{''}$ & $13.5^{d}$ & $0.1^{d}$ & 2857022 & 39.9\\
GX 301-2 & 2733, 3433 & TE & $12^{h}26^{m}38{s}$ & $-62^{d}46^{'}13^{''}$ & $300.1^{d}$ & $0.0^{d}$ & 584812 & 98.2\\
GX 3+1 & 2745 & CC & $17^{h}47^{m}56^{s}$ & $-26^{d}33^{'}49^{''}$ & $2.3^{d}$ & $0.8^{d}$ & 1298798 & 9.2\\
GX 340+0 & 1921 & TE & $16^{h}45^{m}48{s}$ & $-45^{d}36^{'}40^{''}$ & $339.6^{d}$ & $-0.1^{d}$ & 1385351 & 24.0\\
GX 349+2 & 3354, 715 & TE & $17^{h}05^{m}45^{s}$ & $-36^{d}25^{'}23^{''}$ & $349.1^{d}$ & $2.7^{d}$ & 6602027 & 27.0\\
GX 5-1 & 716 & TE & $18^{h}01^{m}08^{s}$ & $-25^{d}04^{'}45^{''}$ & $5.1^{d}$ & $-1.0^{d}$ & 2123066 & 89.1\\
GX 9+1 & 717 & TE & $18^{h}01^{m}32^{s}$ & $-20^{d}31^{'}44^{''}$ & $9.1^{d}$ & $1.2^{d}$ & 1275273 & 89.9 \\
LMC X-1 & 93 & TE & $05^{h}39^{m}40^{s}$ & $-69^{d}44^{'}37^{''}$ & $280.2^{d}$ & $-31.5^{d}$ & 455010 & 21.8\\
Ser X-1 & 700 & TE & $18^{h}39^{m}58^{s}$ & $+05^{d}02^{'}09^{''}$ & $36.1^{d}$ & $4.8^{d}$ & 7171223 & 76.3\\
Vela X-1 & 1927, 1928 & TE & $09^{h}02^{m}07^{s}$ & $-40^{d}33^{'}17^{''}$ & $263.1^{d}$ & $3.9^{d}$ & 1374192 & 62.1\\
Her X-1 & 2749 & TE & $16^{h}57^{m}50^{s}$ & $35^{d}20^{'}33^{''}$ & $58.1^{d}$ & $37.5^{d}$ & & \\
PKS 2155-304 & 1014,1705,3167 & TE & $21^{h}58^{m}52^{s}$ & $-30^{d}13^{'}32^{''}$ & $17.7^{d}$ & $-52.2^{d}$ & &\\
 &333,3706,3708 & & & & & \\
 \enddata
\end{deluxetable}

\subsection{Halo reconstruction}
Although the zeroth order data in CC-mode and the first order data in TE-mode have
either no or less serious pileup, the zeroth order data in CC-mode have only one
dimensional images and the first order image in TE-mode are mixed with photons of
different energies and radii, from which we cannot get the halo's radial profile
directly. Making use of the assumption that the real halo could be an isotropic image,
we have reported the reconstruction of the images of X-ray halos from the data obtained
with the HETGS and/or in CC-mode. The method has been described in detail in Paper I and
Paper II. Here, we briefly review the method and process of the halo reconstruction.
\begin{figure*}
\includegraphics[scale=0.8]{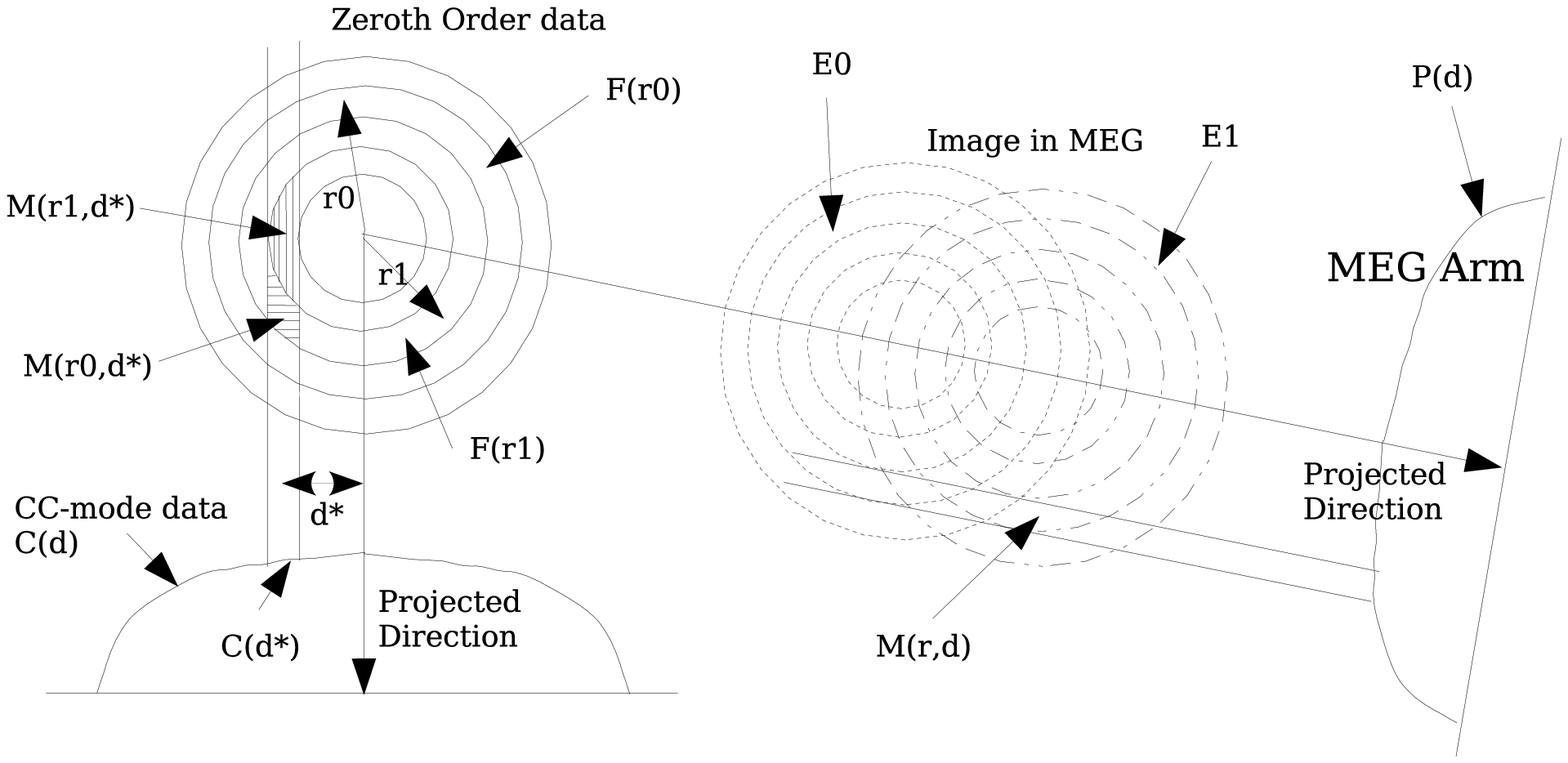}
\caption{The projection of the photons in zeroth order image along the read-out
direction and the projection of the photons along a grating arm.}
\end{figure*}

If the flux of a point source plus its X-ray halo is isotropically distributed and
centered at the point source as $F(r)$, and the projection process, in which the two
dimensional halo image is projected to one dimensional image, can be represented by a
matrix operator M(r, d), then the projected flux distribution P(d) is
\begin{equation}
P(d) = M(r,d) \times F(r),
\end{equation}
where $r$ is the distance from the centroid source position and $d$ is the distance
from the projection center (refer to Fig. 1. cited from Paper II). In CC-mode, we can
only get the count rate $C(d)$, but not the flux projection $P(d)$ directly. With the
exposure map of CCDs calculated, we can get another equation
\begin{equation}
{\rm{exposure}\ \rm{map} \choose \rm{matrix}} \times M(r,d)\times
F(r) = M'(r,d) \times F(r) = C(d),
\end{equation}
where $M'(r,d)={\rm{exposure}\ \rm{map} \choose \rm{matrix}} \times M(r,d)$ is another
matrix. Using the steepest descent method (Marcos $\&$ Benar 2001), we can
solve equation 1. The iterative process can be expressed as
\begin{eqnarray}
F^{(k+1)}=F^{(k)}&+&\frac{[P-MF^{(k)}]^T[P-MF^{(k)}]}{[M(P-MF^{(k)})]^T[P-MF^{(k)}]}
\times [P-MF^{(k)}],
\end{eqnarray}
where $F^{(k+1)}$ and $F^{(k)}$ are the values of F(r) in the ($k+1$)th and $k$th
iterative loops, $[P-MF^{(k)}]^{T}$ is the transpose of the matrix $P-MF^{(k)}$. In our
iterative process, the loop is stopped when
$\frac{1}{N}\sum_{d=1}^{N}(\frac{MF-P}{\Delta P})^2 < 0.05$ or
$\frac{1}{N}\sum_{d=1}^{N}(\frac{MF-P}{\Delta P})^2$ reaches its minimal value, where
$N$ is number of $P(d)$ and $\Delta P$ is the error of $P(d)$. Replacing $M$ in
equation 3 with $M'$, as well as $P$ with $C$, we can solve equation 2.

The accurate {\it Chandra} PSF is important for reconstructing the halo accurately. Following
the same as procedure described in Paper II, we calculate the zeroth order PSF from the
observation data of bright point sources without halo (for example Her X-1 and PKS
2155-304) at large angles and from the MARX\footnote{http://space.mit.edu/ASC/MARX/}
simulation at small angles. Because the data of Her X-1 and PKS 2155-304 in HETGs are
not suffered from pileup even when the source is pointed on-axis, the projection of
grating arms PSF in small angles is also calculated from the observation data. In order
to improve the statistical quality, we use data from seven observations to generate the
PSF. These data IDs are listed in Table 1.

For the TE-mode data, we use the first order data (HEG $\pm 1$ and MEG $\pm 1$) within
60 arcsec around the source position. First, we divided the dataset into 20 energy
bands, spaced every 0.2 keV from 1.0-5.0 keV, and use the process described in Paper II
to generate the pure halo projections. Then we sum them from each energy bands and
derive the halo radial profile in units of photons cm$^{-2}$ s$^{-1}$ arcsec$^{-2}$.
For zeroth order of the CC-mode data, we also use the same process of data reduction
as described in
Paper II to obtain the pure halo radial flux in each energy band. Then we sum the halo
radial flux from each energy band and get the total halo radial flux from 1.0-5.0 keV.

\subsection{Halo model fittings}

As discussed by Mathis $\&$ Lee (1991), the geometry of the X-ray scattering can be
demonstrated as Figure 2. Because all scattering angles are small, $\theta
\approx(1-x)\theta_{sca}$ is always a good approximation, where $x=d/D$, $d$
is the distance of the scattering grain from the observer, and $D$ is the distance to
the source.
\begin{figure*}
\includegraphics[scale=0.8]{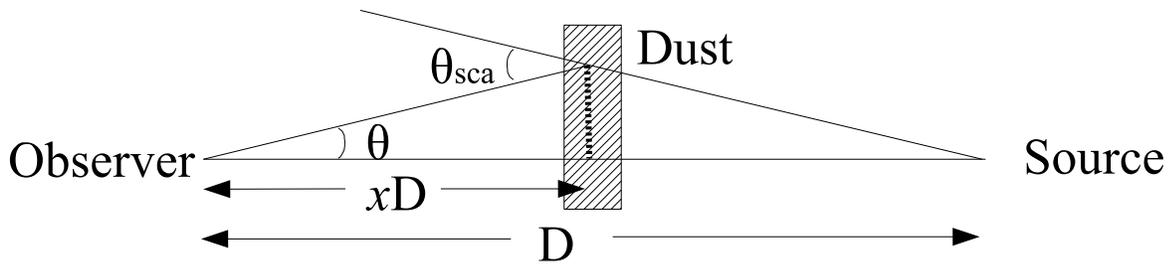}
\caption{Geometry of the X-ray scattering process.  }
\end{figure*}

The observed halo surface brightness $I_{sca}^{(1)}(\theta)$ at an angle $\theta$ from
the point source of X-rays can be shown as
\begin{equation}
{I_{sca}^{(1)} \over F_{X}} = N_{H}\int_{E_{0}}^{E_{1}}{dES(E)}\int_{a_{min}}^{a_{max}}{dan(a)}{\int_{0}^{1}{dxf(x)(1-x)^{-2}}\times {{d\sigma_{sca}(a,E,\theta,x)} \over {d\Omega}}},
\end{equation}
where $N_{H}$ is the total column density of hydrogen between the observer and the
X-rays source, $n(a)da$ is the number of grains per H atom with radii between $a$ and
$a+da$, $F_{X}$ is the observed flux, $S(E)dE$ is the normalized photon energy
distribution of the X-ray point source, $f(x)$ is the density of hydrogen at $xD$
relative to the average density along the line of sight to the X-ray source, and
$d\sigma_{sca} \over {d\Omega}$ is the differential scattering cross section for a
single grain.

Since the X-ray halo is determined by the size, position and composition of the dust
grains, as well as the source flux and absorption column, we need to make some assumptions
about the dust grains before we can use equation 4 to fit the halo profiles. We used the grain
models Weingartner \& Draine (2001, WD01) and Mathis, Rumpl $\&$ Nordsieck (1977, MRN)
to fit our halo radial profile of these sources. The MRN model is a classical model
with the size distributions of dust grains: $n(a)\propto a^{-3.5},\ (0.005\ \mu
m<a<0.25\ \mu m)$. The size distribution of dust grains in WD01 model is rather
complicated, which includes sufficient
very small carbonaceous grains and larger grains (please refer to Weingartner \& Draine 2001 for detail).
Since we only concern the halo in small angles ($< 60^{''}$,
except for the source LMC X-1), multiple scatterings can be neglected even though the
scattering optical depth $\tau_{sca}$ is as large as 2. The model fitting codes were
provided by Randall K. Smith and the same codes have been applied to the halos of
GX 13+1 (Smith, Edgar $\&$ Shafer 2002).

As a first try, we use the smoothly distributed dust models to fit the halo. We find
that these distributions cannot describe the whole halo profile of any source we
analyzed, -- all the $\chi^{2}_{\nu}$ are greater than 5.0 for both the WD01 and MRN
models. Then we used the ``extended" MRN model (Witt, Smith $\&$ Dwek 2001) which has
the same total mass as the MRN model, but extends the MRN size distributions to 2.0
$\mu$m to fit the halo radial profile. We find that these models underestimate the
photons in large angles ($>$20 arcsec) and in very small angles (about 1~3 arcsec), and
overestimate the photons in 3~20 arcsec. The same is true for the "extended" WD01 model
which extends to the size distributions to 2.0 to 5.0 $\mu$m. We then divide the LOS to
four parts ($x$:0.0-0.25, 0.25-0.5, 0.5-0.75, 0.75-1.0), in order to probe the
inhomogeneity at large scales, and assume that the dust grains are smoothly distributed
in each part and that the amount of dust in each part is allowed to vary independently.
However we found that there are systematic excess halo intensity in small angles (below
10 arcsec), indicating that there should be substantial dust grains at the position
very close to the X-ray sources. We also found that the fitting are not very sensitive
in constraining the relative amount of dust grains located at 0.0-0.25 and 0.25-0.50,
but are very sensitive in the one located at 0.75-1.0. We therefore re-divide the LOS
to logarithm-equally spaced bins with more bins close to the sources: 0.0-0.651,
0.651-0.881, 0.881-0.962, 0.962-0.99 and 0.99-1.0; in this scheme the amount of dust grains
in each part may be determined relatively independently. Last we try to use the
power-law or exponential function to fit the dust spatial distributions and find the
fit results are very poor. The best fit results and the values of $\chi^{2}$ for model
are listed in Table 2 and Table 3 for the WD01 and MRN models, respectively. The X-ray
halo radial profiles and relative density distributions along the LOS for all sources
are shown in Figure 3. The fractional halo intensity is calculated from each halo from
$1.5^{''}$ to $60{''}$.

The source LMC X-1 is treated differently from other sources, since it is not in our
galaxy. The detailed fitting process is described in section 3. The halo of LMC X-1
from $1.0^{''}$ to $40^{''}$ is obtained from the grating data and that from $40^{''}$
to $180^{''}$ is extracted directly from the zeroth order ACIS image.
\begin{figure*}
\plotone{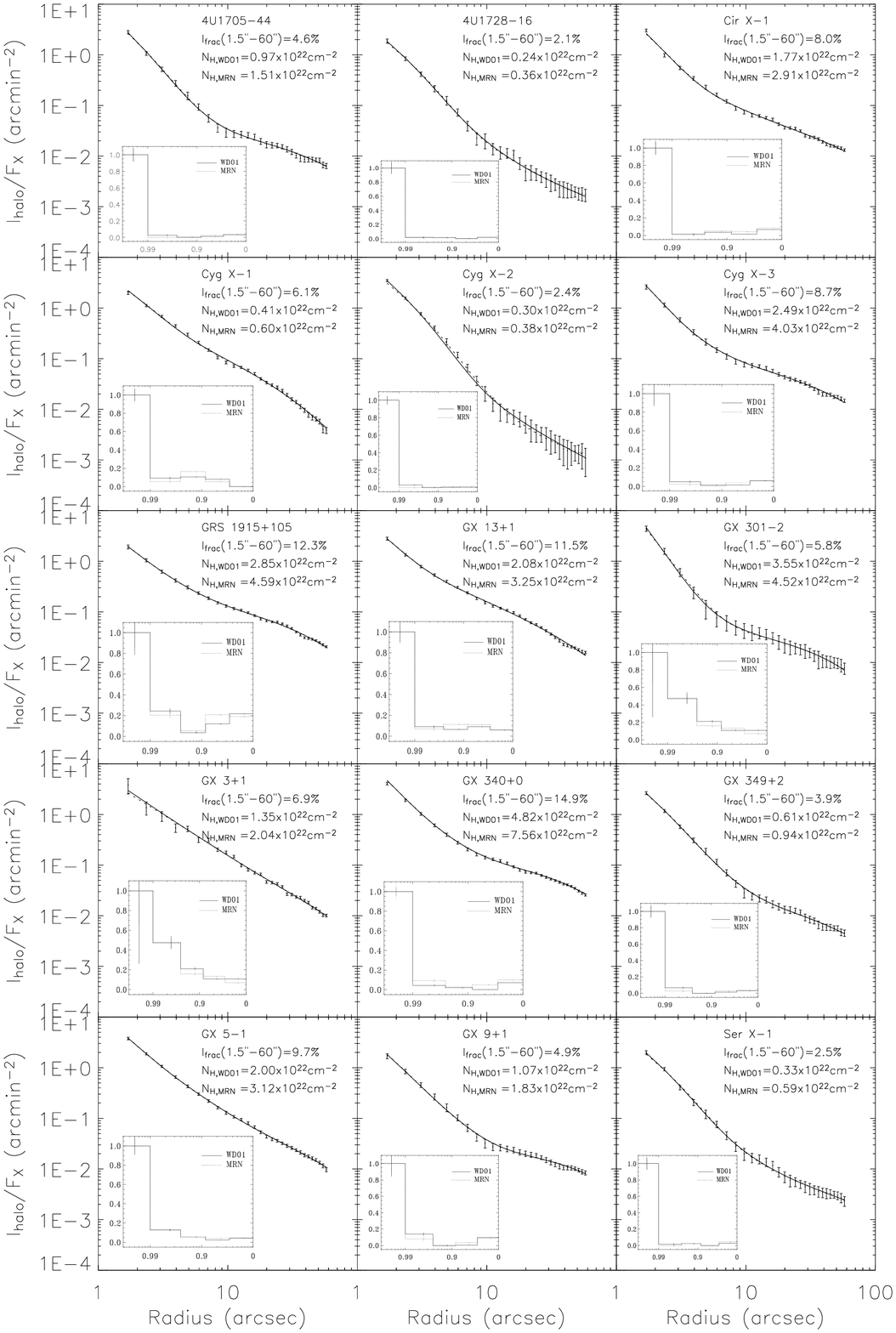}
\end{figure*}
\begin{figure*}
\plottwo{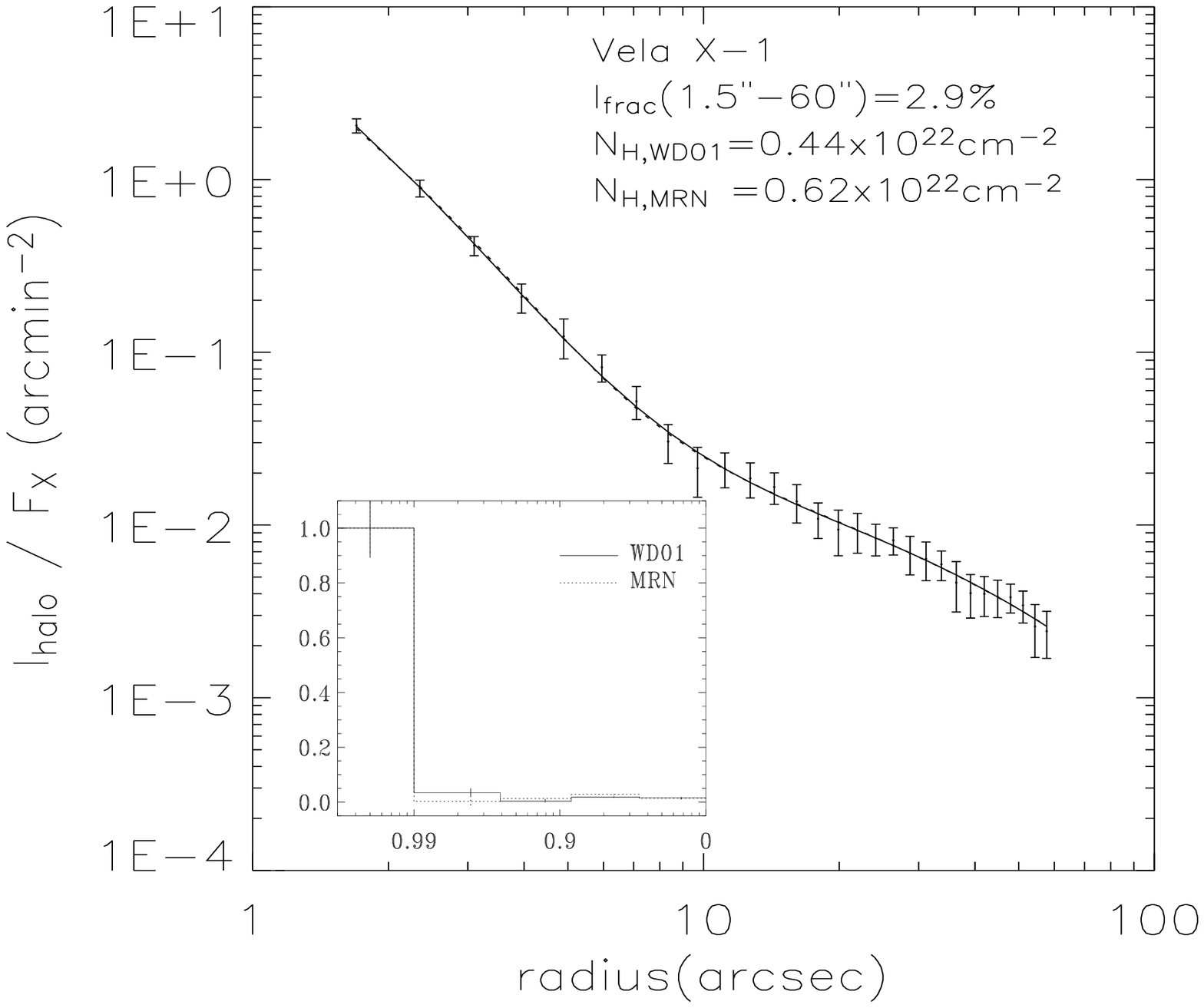}{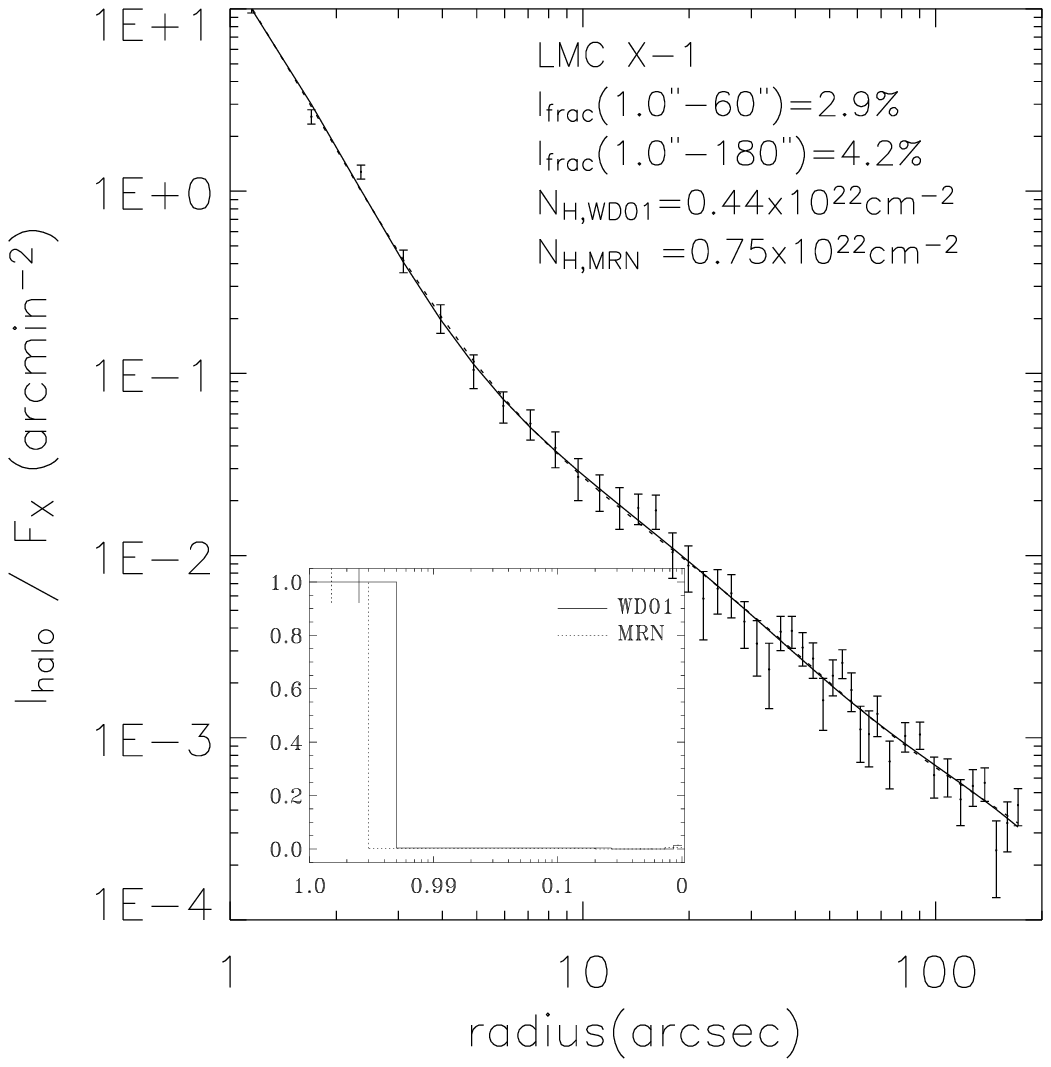} \caption{The X-ray halo profiles of the point X-ray
sources, fitted using the interstellar grain models WD01 and MRN. The solid lines are the model WD01
and the dashed lines are the model MRN. The relative hydrogen density spatial
distributions are shown as insets in the figure, where the horizontal axis is for the
relative distance $x$ to the source and the vertical axis is the relative density
normalized to the value in the last bin (close to 1).}
\end{figure*}

\begin{deluxetable}{rccccccr}
\tabletypesize{\scriptsize} \tablewidth{0pt} \tablecaption{ The relative hydrogen
densities in different segments along the LOS (column 2-6, relative to the peak
hydrogen density along each LOS) and total hydrogen column densities (column 7) from
the grain model WD01 fitting.} \startdata \tableline \tableline
 & \multicolumn{5}{c}{The relative hydrogen density in different position from WD01 fit} & total $N_{H}$ & $\chi^{2}$/dof\\
Source &0.0-0.651 &0.651-0.881 &0.881-0.962 & 0.962-0.99 &0.99-1.0 &$(10^{22}$ cm$^{-2})$ \\
\tableline
4U1705-44 & $0.0331\pm0.0019$  & $0.0118\pm0.0020$ & $0.0000\pm0.0034$ & $0.027\pm0.010$ &$1.00\pm0.08$ & $0.964\pm0.044$ & 13.1/24\\
4U1728-16 & $0.0201\pm0.0031$ & $0.0000\pm0.0029$ & $0.019\pm0.005$ & $0.019\pm0.013$ & $1.00\pm0.08$ & $0.241\pm0.022$ & 4.1/24 \\
Cir X-1 & $0.0672\pm0.0019$ & $0.0138\pm0.0020$ & $0.0359\pm0.0035$ & $0.013\pm0.009$ & $1.00\pm0.08$ & $1.77\pm0.05$  & 32.3/24\\
Cyg X-1 & $0.000\pm0.005$ & $0.0807\pm0.0034$ & $0.106\pm0.005$ & $0.092\pm0.013$ & $1.00\pm0.07$ & $0.413\pm0.021$ & 41.0/24\\
Cyg X-2 & $0.0063\pm0.0019$ & $0.0033\pm0.0017$ & $0.0000\pm0.0031$ & $0.029\pm0.008$ & $1.00\pm0.05$ & $0.295\pm0.027$ & 4.1/24\\
Cyg X-3 & $0.0587\pm0.0020$ & $0.0167\pm0.0024$ & $0.0115\pm0.0043$ & $0.051\pm0.012$ & $1.00\pm0.13$ & $2.49\pm0.09$ & 17.4/24\\
GRS 1915+105 & $0.217\pm0.005$ & $0.119\pm0.005$ & $0.036\pm0.009$ & $0.244\pm0.023$ & $1.00\pm0.21$ & $2.85\pm0.06$ & 33.3/24\\
GX 13+1 & $0.0590\pm0.0025$ & $0.0916\pm0.0029$ & $0.065\pm0.005$ & $0.091\pm0.013$ & $1.00\pm0.11$ & $2.08\pm0.06$ & 44.2/24\\
GX 301-2 & $0.0091\pm0.0009$ & $0.0013\pm0.0012$ & $0.0006\pm0.0022$ & $0.004\pm0.008$ & $1.00\pm0.13$ & $3.55\pm0.33$ & 3.8/24\\
GX 3+1 & $0.107\pm0.005$ & $0.108\pm0.006$ & $0.212\pm0.015$ & $0.48\pm0.07$ & $1.00\pm0.74$ & $1.35\pm0.09$ & 40.9/24\\
GX 340+0 & $0.0710\pm0.0012$ & $0.0029\pm0.0015$ & $0.0250\pm0.0026$ & $0.044\pm0.008$ & $1.00\pm0.14$ & $4.82\pm0.08$ & 19.2/24\\
GX 349+2 & $0.0310\pm0.0023$ & $0.0127\pm0.0022$ & $0.0000\pm0.0041$ & $0.068\pm0.012$ & $1.00\pm0.08$ & $0.611\pm0.032$ & 8.8/24\\
GX 5-1 & $0.0439\pm0.0018$ & $0.0242\pm0.0021$ & $0.0532\pm0.0040$ & $0.128\pm0.011$ & $1.00\pm0.10$ & $2.00\pm0.06$ & 10.5/24\\
GX 9+1 & $0.0948\pm0.0038$ & $0.0041\pm0.0040$ & $0.000\pm0.008$ & $0.138\pm0.020$ &$1.00\pm0.16$ & $1.074\pm0.044$ & 9.9/24\\
Ser X-1 & $0.0278\pm0.0028$ & $0.0000\pm0.0027$ & $0.019\pm0.005$ & $0.015\pm0.012$ & $1.00\pm0.07$ & $0.329\pm0.023$ & 5.2/24\\
Vela X-1 & $0.0158\pm0.0031$ & $0.0186\pm0.0032$ & $0.0037\pm0.006$ & $0.034\pm0.016$ & $1.00\pm0.11$ & $0.436\pm0.041$ & 3.2/24\\
\tableline
 & $x(0.0-1.0):$ & $0.0-0.14_{-0.14}^{+0.45}$ & $0.73_{-0.09}^{+0.06}-0.995$ & $0.995_{-0.0019}^{+0.0012}-1.0$ &  \\
LMC X-1 & relative density & $0.0142\pm0.0017$ & $0.0033\pm0.0003$ & $1.00\pm0.08$ & & $0.444\pm0.027$ & 35.3/41\\
\enddata
\end{deluxetable}

\begin{deluxetable}{rccccccr}
\tabletypesize{\scriptsize}
\tablewidth{0pt}
\tablecaption{The lists of relative hydrogen densities in different positions (column 2-6, relative to the max hydrogen densities in this sight) and total hydrogen column densities (column 7) from the grain model MRN fit.}
\startdata
\tableline \tableline
 & \multicolumn{5}{c}{$N_{H}\ (10^{22}$ cm$^{-2})$ in different position from MRN fit} &total $N_{H}$ & $\chi^{2}$/dof\\
Source &0.0-0.651 &0.651-0.881 &0.881-0.962 & 0.962-0.99 & 0.99-1.0 &$(10^{22}$ cm$^{-2})$ \\
\tableline
4U1705-44 & $0.0382\pm0.0031$  & $0.0276\pm0.0026$ & $0.0052\pm0.0041$ & $0.000\pm0.010$ & $1.00\pm0.07$ & $1.51\pm0.08$ & 13.0/24\\
4U1728-16 & $0.023\pm0.006$ & $0.0098\pm0.0041$ & $0.015\pm0.006$ & $0.014\pm0.012$ & $1.00\pm0.08$ & $0.362\pm0.047$ & 5.7/24\\
Cir X-1 & $0.0838\pm0.0032$ & $0.0408\pm0.0028$ & $0.0495\pm0.0043$ & $0.005\pm0.010$ & $1.00\pm0.07$ & $2.91\pm0.09$ & 33.2/24\\
Cyg X-1 & $0.000\pm0.007$ & $0.0546\pm0.0041$ & $0.165\pm0.006$ & $0.056\pm0.012$ &$1.00\pm0.06$ & $0.596\pm0.048$ & 34.8/24\\
Cyg X-2 & $0.0037\pm0.0031$ & $0.0094\pm0.0022$ & $0.0000\pm0.0030$ & $0.000\pm0.007$ & $1.00\pm0.05$ & $0.38\pm0.06$ & 5.2/24\\
Cyg X-3 & $0.0662\pm0.0031$ & $0.0408\pm0.0030$ & $0.021\pm0.005$ & $0.023\pm0.013$&$1.00\pm0.10$ & $4.03\pm0.15$ & 14.4/24\\
GRS 1915+105 & $0.192\pm0.006$ & $0.207\pm0.006$ & $0.054\pm0.009$ & $0.206\pm0.021$&$1.00\pm0.15$ & $4.59\pm0.11$ & 32.3/24\\
GX 13+1 & $0.0581\pm0.0037$ & $0.1103\pm0.0035$ & $0.114\pm0.006$ & $0.070\pm0.013$&$1.00\pm0.08$ & $3.25\pm0.11$ & 44.0/24\\
GX 301-2 & $0.0138\pm0.0018$ & $0.0067\pm0.0020$ & $0.0000\pm0.0028$ &$0.000\pm0.008$& $1.00\pm0.12$ & $4.47\pm0.37$ & 3.6/24\\
GX 3+1 & $0.070\pm0.006$ & $0.136\pm0.006$ & $0.160\pm0.012$ & $0.468\pm0.049$ & $1.00\pm0.35$ & $2.04\pm0.10$ & 41.2/24\\
GX 340+0 & $0.1015\pm0.0025$ & $0.0511\pm0.0024$ & $0.0147\pm0.0040$ & $0.093\pm0.010$&$1.00\pm0.08$ & $7.56\pm0.16$ & 15.0/24\\
GX 349+2 & $0.0307\pm0.0035$ & $0.0266\pm0.0027$ & $0.0005\pm0.0042$ & $0.028\pm0.011$&$1.00\pm0.06$ & $0.94\pm0.06$ & 9.3/24\\
GX 5-1 & $0.0460\pm0.0027$ & $0.0408\pm0.0026$ & $0.0595\pm0.0045$ &$0.127\pm0.011$& $1.00\pm0.07$ & $3.12\pm0.11$ & 9.6/24\\
GX 9+1 & $0.097\pm0.005$ & $0.0353\pm0.0045$ & $0.000\pm0.008$ & $0.080\pm0.018$&$1.00\pm0.11$ & $1.83\pm0.08$ & 9.8/24\\
Ser X-1 & $0.044\pm0.005$ & $0.0000\pm0.0035$ & $0.025\pm0.005$ & $0.000\pm0.015$&$1.00\pm0.07$ & $0.586\pm0.049$ & 7.8/24\\
Vela X-1 & $0.014\pm0.005$ & $0.0285\pm0.0042$ & $0.013\pm0.007$ & $0.003\pm0.017$&$1.00\pm0.09$ & $0.62\pm0.08$ & 5.5/24\\
\tableline
 & $x(0.0-1.0):$ & $0.0-0.27_{-0.27}^{+0.44}$ & $0.80_{-0.07}^{+0.06}-0.9973$ & $0.9973_{-0.0019}^{+0.0012}-1.0$ & \\
LMC X-1 & relative density & $0.0043\pm0.0006$ & $0.0027\pm0.0003$ & $1.00\pm0.08$ & & $0.75\pm0.05$ & 34.8/41\\
\enddata
\end{deluxetable}

\subsection{Spectral Fittings}
Here we perform spectral fits to derive the hydrogen absorption column densities with
the high resolution spectra from these sources observed with ACIS-S + HETG, in order to
compare with the scattering column densities derived from X-ray halos. We use the
\textit{ciao} 3.0.2 to extract the grating spectra. In order to reduce the
contamination of the halo, the region to filter the data is limited in 5 arcsec for the
TE-mode data.

The HEG$\pm1$ and MEG$\pm1$ spectra sometimes suffered from pileup when the energy is
higher than 2 keV because some sources are very bright, i.e, Cygnus X-1 and Circinus
X-1. However the energy spectra from MEG$\pm3$ are not affected by the pileup and there
are enough photons for spectral analysis. Although the third order photons below
$\sim2$ \AA\ may be contaminated by pileup from the MEG first order for the brightest
source, e.g, Circinus X-1 (Schulz $\&$ Brandt, 2002), this has no impact on our analysis because
we only use the spectra below 5 keV ($\sim 2.4$ \AA). Since there are less photons in
the low energy bands (e.g. less than 2 keV) of MEG$\pm3$ spectra and sometimes the
MEG$\pm$3 spectra are cut off below 2 keV, the HEG$\pm1$ or MEG$\pm1$ energy spectra
below 2 keV band and the MEG$\pm3$ spectra in 2-5 keV bands are also used. The spectra
are rebinned so that there are at least 100 counts in each energy bin for the MEG$\pm$3
spectra, as well 400 counts in each energy bin for the MEG$\pm$1 or HEG$\pm$1 spectra.
We use the Xspec 11.1.0 to fit these energy spectra. Three models, i.e., the power law,
black body and thermal bremsstrahlung, are used, including cold gas absorption, which
was modeled using the absorption cross sections given by Morrison \& McCammon (1983).
For each source, we only list the results for one model which can best fit the data.
The highly accurate emission lines in these spectra are ignored since we only
concern the absorbing hydrogen column density. A summary of our spectral fitting
results are given in Table 4.

\begin{deluxetable}{rrccccc}
\tabletypesize{\scriptsize} \tablewidth{0pt} \tablecaption{List of the spectral fits
(all $N_{H}$ values are in units $10^{22}$ cm$^{-2}$). Our fits used power-law (pwl),
thermal bremsstrahlung (thb) or black body (bb) model.}
\tablehead{ \colhead{Source} &
\colhead{$N_{H}$ (Lit)} & \colhead{$N_{H}$} & \colhead{$\Gamma$ (pwl)} & \colhead{$kT$
(thb)} & \colhead{$kT$ (bb)} & \colhead{$\chi_{\nu}^{2}$/N} } \startdata
4U1705-44 & $1.23^{1}$, $1.45^{2}$, $0.85^{8}$ & $1.23\pm0.04$ & .. & .. & $1.01\pm0.02$ & 0.86/1551\\
4U1728-16 & $0.26^{1}$, $0.26^{2}$, $0.21^{8}$ & $0.24\pm0.02$ & $-1.76\pm0.04$ & .. & .. & 0.95/1547\\
Cir X-1 & $2.41^{1}$, $1.98^{8}$ & $2.41\pm0.05$ & .. & $8.52_{-1.02}^{+1.34}$ & .. & 0.92/574\\
Cyg X-1 & $0.41^{1}$, $0.62^{3}$, $0.72^{8}$ & $0.441\pm0.017$ & $-1.40\pm0.03$ & .. & .. & 1.61/2029\\
Cyg X-2 & $0.25^{1}$, $0.28^{2}$, $0.22^{8}$ & $0.217\pm0.035$ & $-1.43\pm0.05$ & .. & .. & 1.27/382\\
Cyg X-3 & $3.31^{1}$, $1.42^{8}$ & $3.20\pm0.10$ & .. & .. & $1.51\pm0.06$ & 1.38/713\\
GRS 1915+105 & $3.5^{4}$, $1.76^{8}$ & $3.852\pm0.034$ & .. & .. & $0.900\pm0.010$ & 0.86/3169\\
GX 13+1 & $2.79^{1}$, $1.79^{8}$ & $2.671\pm0.027$ & .. & .. & $1.00_{-0.01}^{+0.02}$ & 0.92/2654\\
GX 301-2 & $85^{5}$, $1.81^{8}$ & $16.2\pm0.8$& $0.13\pm0.12$ & .. & .. & 0.82/605\\
GX 3+1 & $1.42^{1}$, $1.2^{8}$ & $1.56\pm0.06$ & .. & .. & $0.72\pm0.05$ & 1.05/926 \\
GX 340+0 & $5.0^{7}$, $2.18^{8}$ & $6.65\pm0.14$ & .. & $13.0_{-2.7}^{+5.4}$ & .. &  1.10/380\\
GX 349+2 & $0.96^{1}$, $0.66^{8}$ & $0.929\pm0.013$ & $1.44\pm0.03$ & .. & .. & 0.96/6080\\
GX 5-1 & $2.78^{1}$,$3.0^{2}$, $0.94^{8}$ & $2.87\pm0.05$ & .. & .. & $1.020\pm0.020$ & 0.85/660\\
GX 9+1 & $1.87^{1}$, $1.43^{1}$, $0.91^{8}$ & $1.339\pm0.040$ & .. & $1.00\pm0.03$ & .. & 0.80/456\\
LMC X-1 & $0.94^{1}$, $0.07^{8}$ & $0.653\pm0.013$ & .. & $2.93\pm0.05$ & .. & 1.41/801\\
Ser X-1 & $0.50^{1}$, $0.47^{8}$ & $0.487\pm0.017$ & .. & $5.75_{-0.19}^{0.22}$ & .. & 1.27/2744\\
Vela X-1 & $1-18^{6}$, $0.51^{8}$ & $1.05\pm0.05$ & $0.98\pm0.08$ & .. & . & 0.99/412\\
\enddata
\tablenotetext{a}{The literature values (column 2) are taken from Predehl $\&$ Schmitt
1995 (1), Schulz 1999 (2), Schulz \emph{et al} 2002b (3), Chapuis $\&$ Corbel 2004 (4),
Watanabe \emph{et al} 2003 (5), Kretschmar \emph{et al} 1997 (6), Gilfanov \emph{et al}
2003 (7) and the galactic H I column density (8) from heasarc website
``http://heasarc.gsfc.nasa.gov/cgi-bin/Tools/w3nh/w3nh.pl".}
\end{deluxetable}

\section{Discussion of results}
The main quantities derived from our X-ray analysis for each of our sample sources are
the equivalent hydrogen absorption column density and the hydrogen scattering column
density $N_{H}$, the spatial distribution of the dust medium and the fractional halo
intensity $I_{frac}$($1.5^{''}< \theta <60^{''}$). In the following we will discuss these
results and investigate the correlations between these measurements.

\subsection{Spatial distribution of dust medium}
In section 2, we have shown that we can use our method to resolve the halos
as close as 1.5 arcsec to the surrounded point sources.
These small angle halos could provide tight constraints on the
different interstellar grain models and on the dust spatial distribution
especially when the dust grains are very close to the sources.

Both the WD01 model and MRN model fits indicate that there should be substantial amount
of dust grains near the sources; these high density dust may be the molecular clouds
in which the binary systems were formed initially (see section 3.5 for more
discussions).

Next we move on to discuss the results of some sources which are either different from
the majority of other sources, or our results are not exactly the same as the
existing results in the literature.

\subsubsection {LMC X-1}
LMC X-1 is located in the Large Magellanic Cloud (LMC), about 55 kpc away from us and
its galactic coordinates are $l=280.2^{d}$, $b=-31.5^{d}$. Predehl $\&$ Schmitt (1995)
observed it with \textit{ROSAT}, and found $f_{halo} = 2.9\%$ at 1.26 keV, less than our result,
$f_{halo} = 4.2\%$ from 1.5 arcsec to 180 arcsec at 1.79 keV, if $f_{halo}$ is
proportional to $E^{-2}$. Predehl $\&$ Schmitt (1995) estimated that the halo and PSF
have equal intensities at about $200^{''}$ for the LMC X-1 observation. But according
to our analysis, the halo intensity is concentrated mostly in small angles:
$f_{halo}(1.0^{''} < \theta < 60^{''}) = 2.9\%$ and $f_{halo}(60^{''}< \theta <
180^{''})=1.3\%$. Thus it appears possible that the much coarse angular resolution of
the \textit{ROSAT} has resulted in under-estimation of total halo intensity.

Since the LMC X-1 is at high galactic latitude and the X-ray photon travels through the two galaxies, we divided the LOS to three parts: the first one is in our galaxy ($0\sim x_{1}$), the second one is in the LMC ($x_{2}\sim1$) and the last one is nearby the source ($x_{3}\sim1$), where $x_{1}$, $x_{2}$, $x_{3}$ are the positions relative to the distance from LMC X-1. We assume the dust grains in each part are smoothly distributed along the LOS. First we change the values of $x_{1}$, $x_{2}$, $x_{3}$ manually to fit the halo profile with the grain model WD01 and get the value of $\chi^{2}$ for each trial. Searching for the minimal $\chi^{2}$, we get the best fit values of $x_{1,2,3}$. Then we separately change the value of $x_{1}$, $x_{2}$, $x_{3}$ and get their uncertainties. The best values are: $x_{1}=0.14_{-0.14}^{+0.45}$, $x_{2}=0.73_{-0.09}^{+0.06}$, $x_{3}=0.995_{-0.0019}^{+0.0012}$. Finally, from the model fits, we get the $N_{H}$ ($10^{22}$ cm$^{-2}$): $N_{H,WD01}^{0-0.14} = 0.112$, $N_{H,WD01}^{0.73-0.995} = 0.050$, $N_{H,WD01}^{0.995-1} = 0.282$. The total hydrogen column density is $0.444 \times 10^{22}$ cm$^{-2}$ and is consistent with $N_{H} = 0.46 \times 10^{22}$ cm$^{-2}$ derived from spectral fitting by Cui \emph{et al} (2002). For the MRN model, the best-fit parameters are similar and listed in table 3.

These results indicate that the dust grains in the high galactic latitude are much less
than the ones nearby the LMC X-1 (within about 0.3 kpc of the source), and the dust
grain density along the LOS (excluding the vicinity of LMC X-1) to the LMC is very low.
It is consistent with the fact that most dust grains are located in the galactic disk.
In fact the $N_{H}$ of LMC X-1 inside the Milky Way derived in our model fittings is in
good agreement with $N_{H}=0.07 \times 10^{22}$ cm$^{-2}$, inferred from the 21-cm
survey as listed in Table 4 (indicated by literature (8)), which is the HI column
density inside the Milky Way but averaged along the approximate direction of LMC X-1
(the closest direction with available data is about 0.3 degree apart from LMC X-1); the
small difference may be caused by the inhomogeneity of the molecular clouds along the
approximate direction of LMC X-1. We mention in passing that it is not meaningful to
make comparisons between $N_{H}$ indicated by literature (8) in Table 4 with $N_{H}$
derived from either the grain model fittings or spectral fittings for all galactic
sources, because the former is obtained by integrating the HI density through the whole
Galaxy along the approximate direction of the particular source and is thus more
adequate for estimating $N_{H}$ of extra-galactic sources.

\subsubsection{Cygnus X-1 and Cygnus X-3}
Cygnus X-1 is the first black hole candidate, located at $l=71.3^{d}$, $b=3.1^{d}$. It
is one of the brightest X-ray sources and the estimated distance is $D = 2.5 \pm 0.3$ kpc (Ninkov, Walker $\&$ Yang, 1987). Schulz \emph{et al} (2002) used the X-ray
absorption spectroscopy observed with {\it Chandra} ACIS-S + HETG to derive the
hydrogen column density, $N_{H}=6.2 \times 10^{21}$ cm$^{-2}$. We have derived the
$f_{halo}(1.5^{''} < \theta < 60^{''}) = 6.1\%$ at 1.99 keV; this is slightly more than
the value $f_{halo} = 11.6\%$ at 1.20 keV, found by Predehl $\&$ Schmitt (1995), but
consistent with the result in Paper I and Paper II.

The grain models WD01 and MRN both show that the dust grains between the observer and
Cygnus X-1 are not smoothly distributed along the LOS and the dust grains very close to
the sources. This is consistent with the result in the Paper II. However the details of
the dust spatial distribution derived here is not the same as the result in Paper II,
because in Paper II we assumed only one segment of scattering dust grains located along
the LOS.

Cygnus X-3 is an X-ray binary with more than 40$\%$ halo flux in 0.1--2.4 keV (\textit{ROSAT}
energy range) (Predehl $\&$ Schmitt, 1995), located at $l=79.8^{d}$, $b=0.7^{d}$. The
bright X-ray halo has been used to determine the distance to the source as about 9 kpc
(Predehl, \emph{et al}, 2000). The halo fraction $f_{halo}(1.5^{''} < \theta <
60^{''})=8.5\%$ in 1.0--5.0 keV is much less than the total halo fraction of \textit{ROSAT}
which includes the large angle halo, indicating a substantial amount of scattering
dust is located at the near side to the observer.

\subsubsection{GX 13+1}
GX 13+1, a LMXB, is a bright and highly absorbed X-ray source and located at
$l=15.5^{d}$, $b=0.1^{d}$. Smith, Edgar $\&$ Shafer (2002) observed the GX 13+1 with
{\it Chandra} ACIS-I, and extracted the halo between $50^{''}$ and $600^{''}$. They
found the halo fraction is $f_{halo}(50^{''} < \theta < 600^{''})\approx12\%$ in 2.8
keV and the total halo intensity as a function of energy is $I(E) =
1.5^{+0.5}_{-0.1}E^{-2}_{keV}$ and $I(E, 50^{''} < \theta < 600^{''}) =
(0.939\pm0.028)E^{-2}_{keV}$. This corresponds to $f_{halo}(0^{''}<\theta < 50^{''}) =
(7.02\sim13.54)\%$ at 2.8 keV. This is consistent with our result
$f_{halo}(1.5^{''}<\theta < 60^{''}) = 11.4\%$ at 2.82 keV.

Our fits show that the total hydrogen column density $N_{H}=2.08\times 10^{22}$
cm$^{-2}$ from the grain model WD01 is significantly larger than the value $1.74 \times
10^{22}$ cm$^{-2}$ derived from the WD01 model fit by Smith, Edgar $\&$ Shafer (2002).
It is possible that Smith, Edgar $\&$ Shafer (2002) underestimated the dust located at
$x>0.90$ which does not affect the observed $I(\theta_{h}>50^{''})$, since these dust
near the source contributes to the halo mainly in angles smaller than $50^{''}$. This
interpretation is also consistent with the hydrogen scattering column density
$N_{H}=1.64\pm0.07 \times 10^{22}$ cm$^{-2}$ derived from the WD01 fitting by
integrating the hydrogen density in $x=0.0-0.90$ from the results in Table 2.

\subsubsection{GX 301-2 and Vela X-1}
The X-ray binary pulsar GX 301-2 is located at $l=300.1^{d}$,
$b=0.0^{d}$. The system consists of an accreting magnetized
neutron star in a highly eccentric orbit, embedded in the stellar
wind from a B2 supergiant companion star. When the orbital phase
is in the pre-periastron, the hydrogen column density can be as large
as $1\times10^{24}$ cm$^{-2}$ (Watanabe \emph{et al}. 2003).

The hydrogen column density from the grain model fit is much smaller than the one from
the spectral fit. A likely interpretation is that there are strong and dense stellar
winds around the X-ray source and they absorb and scatters the X-ray photons severely.
Therefore the $N_{H}$ from the spectral fits reflects the total absorption column
density correctly. However the scattered photons in the halo caused by the stellar
winds are very close to the point source as $x$ is very close to 1, e.g., $\theta <
0.5^{''}$ for $r_{wind}<0.1$ pc which corresponds to $x>0.9999$, where $r_{wind}$ is the
radius of the dense stellar winds. Consequently this part of halo is indistinguishable
from the point source, resulting in underestimation of the scattering column density.

Vela X-1, similar to GX301-2, is an eclipsing high-mass X-ray binary consisting of an
early type supergiant and a pulsar, having an orbital radius of about
$1.7R_{*}\ (\sim53R_{\odot})$ with a period of 8.964 days. The X-ray pulsar is accreting
from the intense stellar winds emanating from the companion and the X-ray spectrum
shows considerable photoelectric absorption at lower energies, varying with the orbital
phase, as the neutron star is deeply embedded in the dense stellar wind (Haberl $\&$
White, 1990). Kretschmar \emph{et al} (1997) found the column density $N_{H}$ varies
from $(1\pm1) \times 10^{22}$ cm$^{-2}$ to $(18\pm4)\times10^{22}$ cm$^{-2}$. Since the
halo caused by the stellar winds has an angular size smaller than $0.5^{''}$, the
scattering column density is also underestimated.

For several other sources, e.g., 4U 1705-44, Cir X-1 and Ser X-1, the halo intensity
between $1.5^{''}$ and $60^{''}$ is also smaller than the total halo intensity obtained
by Predehl $\&$ Schmitt (1995), most likely due to the smaller angular size studied
here.

\subsection{Correlations between different $N_{H}$ column densities}

In Figure 4 we show the correlation between $N_{H,MRN}$ and $N_{H,WD01}$ which is well
fitted by a linear relation, $N_{H, MRN}=(1.594\pm0.023)\times
N_{H,WD01}+(-0.009\pm0.025)$. This shows that the column density derived by fitting the
X-ray halo using the MRN model is $\sim60\%$ larger than the fitting results using the
WD01 model. This is close to the result ($\sim45\%$) of Smith, Edgar $\&$ Shafer
(2002). The systematic difference for the inferred $N_{H}$ in the two models is most
likely due to more large-grains in WD01 model than in MRN model, which scatter photons
preferentially in small angles, i.e, more total scattering cross section
$\int_{a_{min}}^{a_{max}} a^4dn(a)$ of the dust per H nucleus in the WD01 model, about
65$\%$ more than in the MRN model.

\begin{figure}
\includegraphics[scale=0.75]{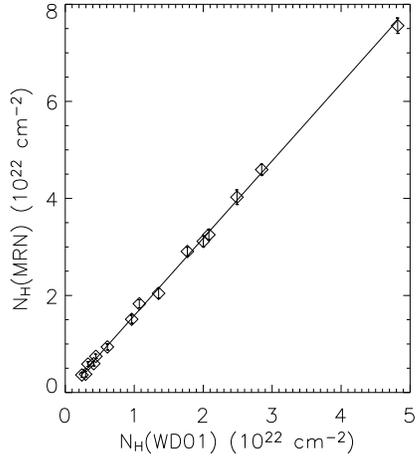}
\caption{Correlation between the scattering column density $N_{H}$ derived by fitting
the X-ray halos with MRN and WD01 models respectively.}
\end{figure}

In Figure 5 we show the correlations between the dust scattering column densities
$N_{H, WD01}$ and $N_{H, MRN}$ derived from the two models, and X-ray absorbing column
density $N_{H,abs}$ derived from X-ray spectral fitting. It is obvious from Figure 5
that both correlations are extremely tight, which can be described by two linear
relations $N_{H,WD01} = (0.720\pm0.009) \times N_{H,abs} + (0.051\pm0.013)$ and $N_{N, MRN} =
(1.156\pm0.016) \times N_{H,abs} + (0.062\pm0.024)$, in units of $10^{22}$ cm$^{-2}$. The outliers, i.e. GX 301-2 and Vela X-1 as discussed earlier, are not
included in the fitting.

\begin{figure}
\includegraphics{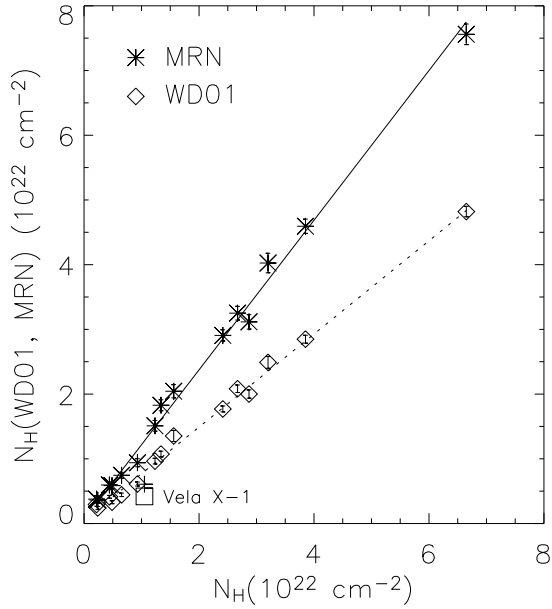}
\caption{ The correlations between the dust scattering column densities $N_{H, WD01}$
and $N_{H, MRN}$ derived from the two models, and X-ray absorbing column density
$N_{H,abs}$ derived from X-ray spectral fitting.}
\end{figure}

\subsection{FHI and $N_{H}$}
The fractional halo intensity (FHI) $I_{frac}$ can be shown as (Mathis $\&$ Lee 1991):
\begin{equation}
I_{frac} = {I_{halo} \over (I_{halo}+F_{X})} = (1-e^{-\tau_{sca}}),
\end{equation}
where the optical depth in scattering $\tau_{sca}$ equals to $\tau_{sca}(E=1keV) \times
(E_{mean}/1keV)^{2}$ within Gaussian approximation. The value for the mean energy
$E_{mean}$ has been obtained by weighting each X-ray photon energy according to the
energy dependence of the scattering cross section, i.e. approximately $E^{-2}$. The
fractional halo intensity for all sources and the mean energy $E_{mean}$ are shown in
Table 5.

\begin{deluxetable}{rrccc}
\tabletypesize{\scriptsize}
\tablewidth{0pt}
\tablecaption{Lists
of the fractional halo intensity (column 2), the mean energy
weighted by the scattering cross section (column 3) and the
hydrogen absorption column densities (column 4).} \tablehead{
\colhead{Source} & \colhead{$I_{frac}(\%)$} &
\colhead{$(E_{mean}/1keV)^{2}$} & \colhead{$N_{H}(10^{22}$ cm$^{-2})$} &
\colhead{${N_{H}}\over{E^{2}}$} } \startdata
4U 1705-44 & 4.6 & 2.53 & $1.23\pm0.04$ & 0.192\\
4U 1728-16 & 2.1 & 1.90 & $0.238\pm0.015$ & 0.066\\
Cir X-1 & 8.0 & 2.64 & $2.414\pm0.048$ & 0.346\\
Cyg X-1 & 6.1 & 1.99 & $0.441\pm0.017$ & 0.111\\
Cyg X-2 & 2.4 & 1.94 & $0.217\pm0.035$ & 0.058\\
Cyg X-3 & 8.7 & 3.20 & $3.20\pm0.10$ & 0.313\\
GRS 1915+105 & 12.3 & 2.94 & $3.852\pm0.034$ & 0.446\\
GX 13+1 & 11.5 & 2.82 & $2.671\pm0.027$ & 0.336\\
GX 301-2 & 5.8 & 4.00 & $16.2\pm0.8$ & 1.012\\
GX 3+1 & 6.9 & 2.46 & $1.56\pm0.06$ & 0.258\\
GX 340+0 & 14.9 & 3.25 & $6.65\pm0.14$ & 0.630\\
GX 349+2 & 3.9 & 2.24 & $0.929\pm0.013$ & 0.185\\
GX 5-1 & 9.7 & 2.90 & $2.87\pm0.05$ & 0.342\\
GX 9+1 & 4.9 & 2.55 & $1.339\pm0.040$ & 0.206\\
LMC X-1 & 2.9 & 1.79 & $0.653\pm0.013$ & 0.204\\
Ser X-1 & 2.5 & 1.95 & $0.487\pm0.017$ & 0.128\\
Vela X-1 & 2.9 & 2.35 & $1.05\pm0.05$ & 0.190\\
\enddata
\end{deluxetable}

\begin{figure}
\includegraphics{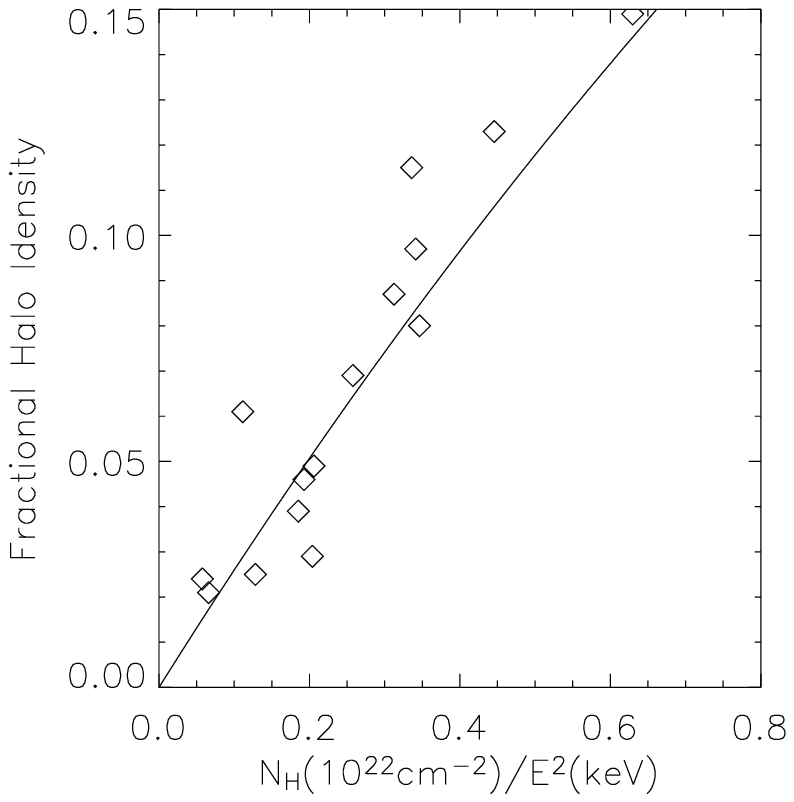}
\caption{The correlation between the fractional halo intensity (FHI)
and the hydrogen column density. The solid line is plotted according to equation 6 with $m=0.53$.}
\end{figure}

Since $\tau_{sca}(E=1keV) \sim N_{H} \times (E/1keV)^{-2}$, equation 5 can be rewritten to
\begin{equation}
I_{frac} = 1-e^{k \times N_{H}[10^{22} cm^{-2}] \over E^{2}},
\end{equation}
where $k$ is approximately a constant. Using the result of Predehl $\&$ Schmitt (1995)
to estimate the value of $k$, we can get $k\approx0.5$.

In our analysis we calculated only the halos between $1.5^{''}$ and $60^{''}$. The
fractional halo intensity $I_{frac}(1.5^{''}<\theta <60^{''})$ will be less than
$I_{frac}(total)$. For each source, $I_{frac}(1.5^{''}<\theta <60^{''}) = m \times
I_{frac}(total)$, $m$ depends on the mean energy, the dust spatial distribution and the
total hydrogen column density of each source, i.e., $m$ varies across the sources.
However in Figure 6, a good correlation is quite obvious, indicating that $m$ does not
change significantly for the sources we have analyzed, excluding the two outliers GX
301-2 and Vela X-1. The solid curve in Figure 6, which best describes the correlation,
is a plot of equation 6 with $m=0.53$.

\subsection{Correlation between $N_{H}$ and the galactic latitude}

If the scale height of dust grains in the disk of Galaxy is $d$ and the X-ray point
source is outside the dust grain layer, the effective distance which X-ray photons
travel through the galactic dust layer is $d/sin(|b^{II}|)$, where $b^{II}$ is the
galactic latitude. Although the dust grains are not smoothly distributed, the
approximately linear correlation between hydrogen scattering column density $N_{H}$ and
the effective distance is expected statistically. Figure 7 shows the correlations with
the typical scale height $d=100$ pc for both MRN and WD01 models, where $N_{H}$ is only
for $x=0.0-0.99$, i.e., $N_{H}$ in $x=0.99-1.0$ is not included since we believe these
scattering media are localized around the sources and should not be correlated with the
distance from the observer. The sources whose galactic latitudes are less than
$0.3^{d}$ are also not included, because $0.1/sin(0.3^{0})>16$ kpc, i.e., these sources
are within the dust layer and the real distance is less than the effective distance of
$d/sin(|b^{II}|)$. We notice that the source GX 3+1 is significantly below the lines.
One possibility is that this source is actually located within the galactic dust layer
and thus its true distance is much less than the effective distance. On the other hand
if the correlations are used as distance indicators for sources in the dust layer, the
fitted $N_{H}$ value would indicate a distance of 4.8 or 5.3 kpc (as indicated in the figure),
which is in remarkable agreement to 4.5 kpc obtained by Kuulkers $\&$ Klis (2000) based
on the Eddington luminosity X-ray bursting property of GX 3+1. As a sanity check, we
also examined the correlations with either the column density in $x=0.99-1.0$ or the
total column density in $x=0.0-1.0$. The former correlation is extremely weak,
consistent with the interpretation the column density in $x=0.99-1.0$ is localized to
each source. The latter correlation, though statistically significant (largely because
the column density in  $x=0.99-1.0$ is much smaller than in  $x=0.0-0.99$ for most
sources), is still much poorer than that shown in Figure 7.
\begin{figure}
\includegraphics[scale=0.6]{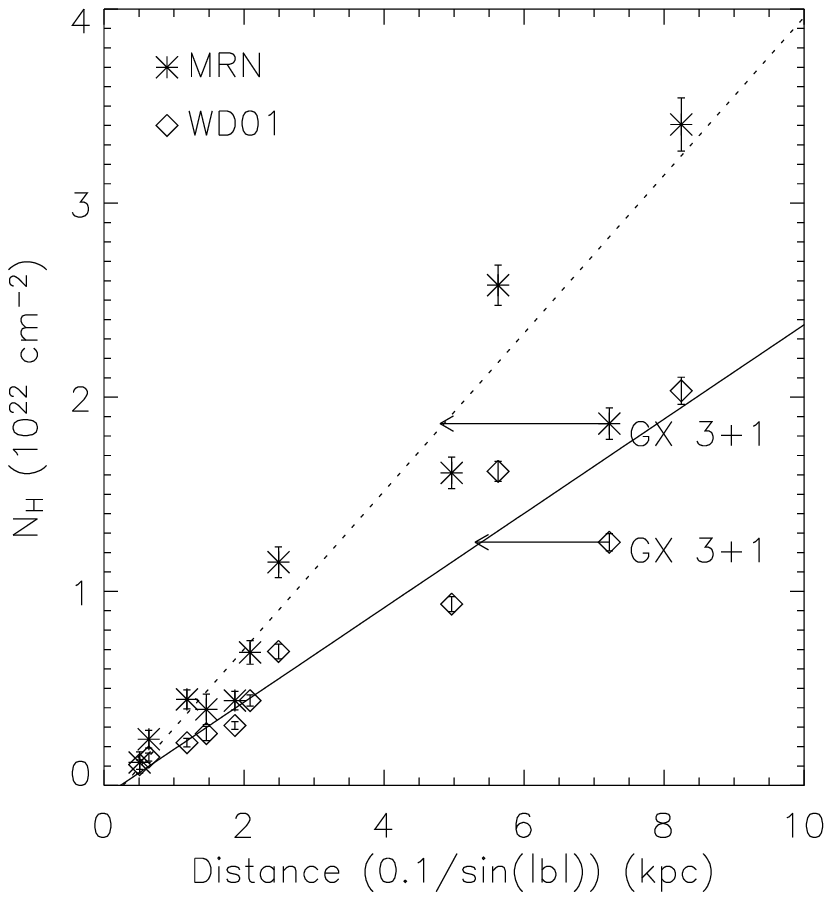}
\caption{The hydrogen scattering column density vs. distance through the galactic dust layer.}
\end{figure}

\subsection{Evidence of molecular clouds around these X-ray binaries?}

In section 2, we have suggested that there should be substantial dust grains near the
sources and these high density dust may be the molecular clouds in which the binary
systems were formed initially. Using the same method for fitting the halo profile of
LMC X-1, it is possible to determine the sizes of these molecular clouds. We use this
method to fit the halo profile of Cygnus X-1 as a test and the best fit parameters are:
$N_{H}=0.147\times10^{22}$ cm$^{-2}$ in 0.994-1.0 for WD01 model and
$N_{H}=0.153\times10^{22}$ cm$^{-2}$ in 0.992-1.0 for MRN model. Assuming the distance
to Cygnus X-1 as 2.5 kpc and Cygnus X-1 is in the center of the assumed molecular
cloud, we can calculate the radius of the cloud, $r_{WD01}=16$ pc, the density,
$\rho_{WD01}\sim30$ cm$^{-3}$, and the total mass, $M_{WD01}\sim1.2\times
10^{4}M_{\odot}$ for the WD01 model. Similarly $r_{MRN}=20$ pc, $\rho_{MRN}\sim25$
cm$^{-3}$ and $M_{MRN}\sim2\times 10^{4}M_{\odot}$ for the MRN model. These results are
consistent with the typical values of molecular clouds.

We therefore suggest that the high angular resolution X-ray halos observed with {\it
Chandra} provide evidence for the existence of molecular clouds surrounding these X-ray
binaries. Unfortunately for most of these sources the data quality, i.e., the limited
accuracy in determining the halos at angles around $1^{''}$, does not allow us to carry
out detailed studies on the exact sizes of these molecular clouds. Future
sub-arcsecond, or even much finer angular resolution X-ray telescopes (e.g., built with
X-ray interferometry technique, Cash {\it et al.} 2000), will allow not only the
determination of the spatial distribution of these molecular clouds, but also
investigations of the supernova remnants of these neutron stars and possibly also black
holes, and even the stellar winds of their companions.

\section{Conclusions}
With excellent angular resolution, good energy resolution and broad energy band, the
{\it Chandra} ACIS is so far the best instrument for studying the X-ray halos. Although
the direct images of bright sources obtained with ACIS usually suffer from severe
pileup which prevents us from obtaining the halos in small angles, we can use our
method to resolve the point source halo from the {\it Chandra} CC-mode data or grating
data and get the halos in very small angles.

In this paper, we analyzed 17 bright X-rays point sources using our method. The halos
between $1.5^{''}$ and $60^{''}$ have been obtained. Then we use the grain models WD01
and MRN to fit the halo radial profiles and derive the hydrogen scattering column
densities and the spatial distributions of dust grains for these sources. Both the WD01
and MRN model fittings indicate that there are substantial dust grains close to the
sources. We suggest this as evidence for the existence of molecular clouds surrounding
these X-ray binaries.

The scattering hydrogen column densities derived from the interstellar grain model WD01 and MRN both have the linear correlations with the absorbing column densities derived from the
spectral model fits (power law model, thermal bremsstrahlung or black body),
$N_{H,WD01} = (0.720\pm0.009) \times N_{H,abs} + (0.051\pm0.013)$ and $N_{N, MRN} =
(1.156\pm0.016) \times N_{H,abs} + (0.062\pm0.024)$, in units of $10^{22}$ cm$^{-2}$.
Both interstellar grain models can describe the observed halo profiles adequately, and it is
difficult to determine which grain model is preferred from our analysis.

At present we cannot obtain dust grain spatial distributions between $x=0.0-0.6$ for
these sources in our Galaxy because the lack of halo intensity distribution covering a
wide range of scattering angles. However the additional data to be collected with {\it
Chandra} in the future for these sources studied here and especially more sources
throughout the galaxy may allow us to probe the spatial distribution of interstellar
dust media in many parts of the Milky Way.

\acknowledgments We thank the anonymous referee for many insightful and constructive
comments and suggestions, which have allowed us to clarified several issues and improve
the presentation of the manuscript significantly. J. Xiang thanks Yuxin Feng and
Xiaoling Zhang for useful discussions and insightful suggestions, and Dr Randall K.
Smith for providing the model codes. This study is supported in part by the Special
Funds for Major State Basic Research Projects and by the National Natural Science
Foundation and the Ministry of Education of China. SNZ also acknowledges supports by
NASA's Marshall Space Flight Center and through NASA's Long Term Space Astrophysics
Program, as well as the {\it Chandra} guest investigation program for the archival
study of X-ray dust scattering halos. YY acknowledges the support by NASA under grant
AR4-5004A.

\end{document}